%% file: prl_magnetic_acceleration.tex
\documentclass[aps,prl, reprint, superscriptaddress]{revtex4-2}

\usepackage{graphicx}	% Including figure files
\usepackage{amsmath}	% Advanced maths commands
\usepackage{amssymb}	% Extra maths symbols
\usepackage{natbib}
\usepackage{color}
\usepackage{amsmath}
\usepackage{subeqnarray}
\usepackage{multirow}
\usepackage{subcaption}

\begin{document}
\title{Magnetic-induced force noise in LISA Pathfinder free-falling test masses}

\input{include/authorlist_september_2019}

%\author{Me}
%\email{mail@example.com}
%\author{Myself}
%\author{Someone Else}
%\affiliation{A University}

\begin{abstract}
LISA Pathfinder was a mission designed to test key technologies required for gravitational wave detection in space. Magnetically driven forces play a key role in the instrument sensitivity in the low-frequency regime, which corresponds to the measurement band of interest for future space-borne gravitational wave observatories. Magnetically-induced forces couple to the test mass motion, introducing a contribution to the relative acceleration noise between the free falling test masses. In this Letter we present the first complete estimate of this term of the instrument performance model. Our results set the magnetic-induced acceleration noise during the February 2017 noise run of $\rm 0.25_{-0.08}^{+0.15}\,fm\,s^{-2}/\sqrt{Hz}$ at 1 mHz and $\rm 1.01_{-0.24}^{+0.73}\, fm\,s^{-2}/\sqrt{Hz}$ at 0.1 mHz. We also discuss how the non-stationarities of the interplanetary magnetic field can affect these values during extreme space weather conditions.
\end{abstract}

\maketitle

%%%%%%%%%%%%%%%%% BODY OF PAPER %%%%%%%%%%%%%%%%%%

\input{include/main}

%%%%%%%%%%%%%%%%%%%%%%%%%%%%%%%%%%%%%%%%%%%%%%%%%%

%%%%%%%%%%%%%%%%% APPENDICES %%%%%%%%%%%%%%%%%%%%%

\appendix

%\input{include/appendix}


%%%%%%%%%%%%%%%%%%%%%%%%%%%%%%%%%%%%%%%%%%%%%%%%%%



\bibliographystyle{apsrev4-2}
\bibliography{library} % if your bibtex file is called example.bib

\end{document}

%% file: include/authorlist_september_2019.tex
%\documentclass[notitlepage,aps,prl,reprint,superscriptaddress,showpacs]{revtex4-1}
%
%\usepackage{graphicx}
%\usepackage{amsmath,amssymb, graphics, setspace}
%\usepackage{commath}
%\usepackage{float}
%\usepackage{color}
%\usepackage[normalem]{ulem}
%\newcommand{\mathsym}[1]{{}}
%\newcommand{\unicode}[1]{{}}
%
%\newcounter{mathematicapage}
%
%
%\begin{document}
%
%
%
%
%\title{
%Author list : 5 September 2019}

\def\addressa{European Space Astronomy Centre, European Space Agency, Villanueva de la
Ca\~{n}ada, 28692 Madrid, Spain}
\def\addressb{Albert-Einstein-Institut, Max-Planck-Institut f\"ur Gravitationsphysik und Leibniz Universit\"at Hannover,
Callinstra{\ss}e 38, 30167 Hannover, Germany}
\def\addressc{APC, Univ Paris Diderot, CNRS/IN2P3, CEA/lrfu, Obs de Paris, Sorbonne Paris Cit\'e, France}
\def\addressd{High Energy Physics Group, Physics Department, Imperial College London, Blackett Laboratory, Prince Consort Road, London, SW7 2BW, UK }
\def\addresse{Dipartimento di Fisica, Universit\`a di Roma ``Tor Vergata'',  and INFN, sezione Roma Tor Vergata, I-00133 Roma, Italy}
\def\addressf{Department of Industrial Engineering, University of Trento, via Sommarive 9, 38123 Trento, 
and Trento Institute for Fundamental Physics and Application / INFN}
\def\addressh{European Space Technology Centre, European Space Agency, 
Keplerlaan 1, 2200 AG Noordwijk, The Netherlands}
\def\addressi{Dipartimento di Fisica, Universit\`a di Trento and Trento Institute for 
Fundamental Physics and Application / INFN, 38123 Povo, Trento, Italy}
\def\addressk{Istituto di Fotonica e Nanotecnologie, CNR-Fondazione Bruno Kessler, I-38123 Povo, Trento, Italy}
\def\addressj{The School of Physics and Astronomy, University of
Birmingham, Birmingham, UK}
\def\addressl{Institut f\"ur Geophysik, ETH Z\"urich, Sonneggstrasse 5, CH-8092, Z\"urich, Switzerland}
\def\addressm{The UK Astronomy Technology Centre, Royal Observatory, Edinburgh, Blackford Hill, Edinburgh, EH9 3HJ, UK}
\def\addressn{Institut de Ci\`encies de l'Espai (ICE, CSIC), Campus UAB, Carrer de Can Magrans s/n, 08193 Cerdanyola del Vall\`es, Spain}
\def\addresso{DISPEA, Universit\`a di Urbino ``Carlo Bo'', Via S. Chiara, 27 61029 Urbino/INFN, Italy}
\def\addressp{European Space Operations Centre, European Space Agency, 64293 Darmstadt, Germany}
\def\addressq{Physik Institut, 
Universit\"at Z\"urich, Winterthurerstrasse 190, CH-8057 Z\"urich, Switzerland}
\def\addressr{SUPA, Institute for Gravitational Research, School of Physics and Astronomy, University of Glasgow, Glasgow, G12 8QQ, UK}
\def\addresss{Department d'Enginyeria Electr\`onica, Universitat Polit\`ecnica de Catalunya,  08034 Barcelona, Spain}
\def\addresst{Institut d'Estudis Espacials de Catalunya (IEEC), C/ Gran Capit\`a 2-4, 08034 Barcelona, Spain}
\def\addressu{Gravitational Astrophysics Lab, NASA Goddard Space Flight Center, 8800 Greenbelt Road, Greenbelt, MD 20771 USA}
\def\addressbb{Department of Mechanical and Aerospace Engineering, MAE-A, P.O. Box 116250, University of Florida, Gainesville, Florida 32611, USA}
\def\addresscc{Istituto di Fotonica e Nanotecnologie, CNR-Fondazione Bruno Kessler, I-38123 Povo, Trento, Italy}
\def\addressdd{isardSAT SL, Marie Curie 8-14, 08042 Barcelona, Catalonia, Spain}
\def\addressee{Escuela Superior de Ingenier\'ia, Universidad de C\'adiz, 11519 C\'adiz, Spain}
\def\addressen{Observatoire de la C\^{o}te d'Azur, Boulevard de l'Observatoire CS 34229 - F 06304 NICE, France}

%  30 September 2012
%  30 March 2017

\author{M~Armano}\affiliation{\addressh}
\author{H~Audley}\affiliation{\addressb}
\author{J~Baird}\affiliation{\addressc}
\author{P~Binetruy}\thanks{Deceased}\affiliation{\addressc}
\author{M~Born}\affiliation{\addressb}
\author{D~Bortoluzzi}\affiliation{\addressf}
\author{E~Castelli}\affiliation{\addressi}
\author{A~Cavalleri}\affiliation{\addresscc}
\author{A~Cesarini}\affiliation{\addresso}
\author{A\,M~Cruise}\affiliation{\addressj}
\author{K~Danzmann}\affiliation{\addressb}
\author{M~de Deus Silva}\affiliation{\addressa}
\author{I~Diepholz}\affiliation{\addressb}
\author{G~Dixon}\affiliation{\addressj}
\author{R~Dolesi}\affiliation{\addressi}
\author{L~Ferraioli}\affiliation{\addressl}
\author{V~Ferroni}\affiliation{\addressi}
\author{E\,D~Fitzsimons}\affiliation{\addressm}
\author{M~Freschi}\affiliation{\addressa}
\author{L~Gesa}\thanks{Deceased}\affiliation{\addressn}\affiliation{\addresst}
\author{F~Gibert}\affiliation{\addressi}\affiliation{\addressdd}
\author{D~Giardini}\affiliation{\addressl}
\author{R~Giusteri}\affiliation{\addressb}
\author{C~Grimani}\affiliation{\addresso}
\author{J~Grzymisch}\affiliation{\addressh}
\author{I~Harrison}\affiliation{\addressp}
\author{M-S~Hartig}\affiliation{\addressb}
\author{G~Heinzel}\affiliation{\addressb}
\author{M~Hewitson}\affiliation{\addressb}
\author{D~Hollington}\affiliation{\addressd}
\author{D~Hoyland}\affiliation{\addressj}
\author{M~Hueller}\affiliation{\addressi}
\author{H~Inchausp\'e}\affiliation{\addressc}\affiliation{\addressbb}
\author{O~Jennrich}\affiliation{\addressh}
\author{P~Jetzer}\affiliation{\addressq}
\author{N~Karnesis}\affiliation{\addressc}
\author{B~Kaune}\affiliation{\addressb}
\author{N~Korsakova}\affiliation{\addressen}
\author{C\,J~Killow}\affiliation{\addressr}
\author{J\,A~Lobo}\thanks{Deceased}\affiliation{\addressn}\affiliation{\addresst}
\author{L~Liu}\affiliation{\addressi}
\author{J\,P~L\'opez-Zaragoza}\email{jplopez@ice.csic.es}\affiliation{\addressn}\affiliation{\addresst}
\author{R~Maarschalkerweerd}\affiliation{\addressp}
\author{D~Mance}\affiliation{\addressl}
\author{N~Meshksar}\affiliation{\addressl}
\author{V~Mart\'{i}n}\affiliation{\addressn}\affiliation{\addresst}
\author{L~Martin-Polo}\affiliation{\addressa}
\author{J~Martino}\affiliation{\addressc}
\author{F~Martin-Porqueras}\affiliation{\addressa}
\author{P\,W~McNamara}\affiliation{\addressh}
\author{J~Mendes}\affiliation{\addressp}
\author{L~Mendes}\affiliation{\addressa}
\author{M~Nofrarias}\email{nofrarias@ice.csic.es}\affiliation{\addressn}\affiliation{\addresst}
\author{S~Paczkowski}\affiliation{\addressb}
\author{M~Perreur-Lloyd}\affiliation{\addressr}
\author{A~Petiteau}\affiliation{\addressc}
\author{P~Pivato}\affiliation{\addressi}
\author{E~Plagnol}\affiliation{\addressc}
\author{J~Ramos-Castro}\affiliation{\addresss}\affiliation{\addresst}
\author{J~Reiche}\affiliation{\addressb}
\author{D\,I~Robertson}\affiliation{\addressr}
\author{F~Rivas}\affiliation{\addressn}\affiliation{\addresst}
\author{G~Russano}\affiliation{\addressi}
\author{L~Sala}\affiliation{\addressi}
\author{D~Serrano}\email{dserrano@ice.csic.es}\affiliation{\addressn}\affiliation{\addresst}
\author{J~Slutsky}\affiliation{\addressu}
\author{C\,F~Sopuerta}\affiliation{\addressn}\affiliation{\addresst}
\author{T~Sumner}\affiliation{\addressd}
\author{D~Texier}\affiliation{\addressa}
\author{J\,I~Thorpe}\affiliation{\addressu}
\author{D~Vetrugno}\affiliation{\addressi}
\author{S~Vitale}\affiliation{\addressi}
\author{G~Wanner}\affiliation{\addressb}
\author{H~Ward}\affiliation{\addressr}
\author{P\,J~Wass}\affiliation{\addressd}\affiliation{\addressbb}
\author{W\,J~Weber}\affiliation{\addressi}
\author{L~Wissel}\affiliation{\addressb}
\author{A~Wittchen}\affiliation{\addressb}
\author{P~Zweifel}\affiliation{\addressl}

%
%\date{draft of \today}
%
%
%
%
%\maketitle
%
%
%\end{document}

%% file: include/main.tex
%% INTRO & DESCRIPTION

\paragraph{Introduction} 

LISA Pathfinder~\citep{Anza05, Antonucci12} was an ESA mission with NASA contributions designed to test key technologies for the future gravitational-wave observatory in space, the Laser Interferometry Space Antenna (LISA)~\citep{Amaro17}. LISA Pathfinder was launched on December 3rd 2015 and, after a  one-month cruise phase, reached the L1 Lissajous orbit where it operated until July 2017.
The main scientific result of the mission was to demonstrate the level of relative acceleration between free falling test masses (TMs) required to detect gravitational wave from space~\citep{Armano16, Armano18}. 
The main experiment on-board the LISA Pathfinder mission consisted of two gold-platinum test masses sitting within a 6 degree-of-freedom capacitive position sensor and actuator, the Gravitational Reference Sensor (GRS) ~\citep{Dolesi03, Armano17_capacitive}. 
The position and orientation of the test masses was continuously monitored by a high precision interferometric readout system, the Optical Metrology System (OMS)~\citep{Armano21, Armano22}. These measurements were fed to the on-board control system in charge of isolating the test mass from the perturbations coming from outer space ~\citep{Armano19_dragfree}. This architecture, the so-called drag free control loop, ensured the required free fall purity of one of the test masses by means of a continuous actuation on the spacecraft attitude through the micro-Newton propulsion system~\citep{Armano19_thrusters}. 
The drag free control loop was not intended to isolate the test mass free fall from forces arising internally to the spacecraft. Instead, mission and payload were designed to ensure that these effects did not affect the instrument performance. This had implications, for instance, on the selected orbit  ---the L1 orbit allowed a very thermally stable environment as well as reduced gravitational field gradients--- or several design drivers in the payload to isolate the test masses from potential disturbing forces, typically of thermal or magnetic origin. 

\begin{figure}[t]
\begin{center}
\includegraphics[width=0.5\textwidth]{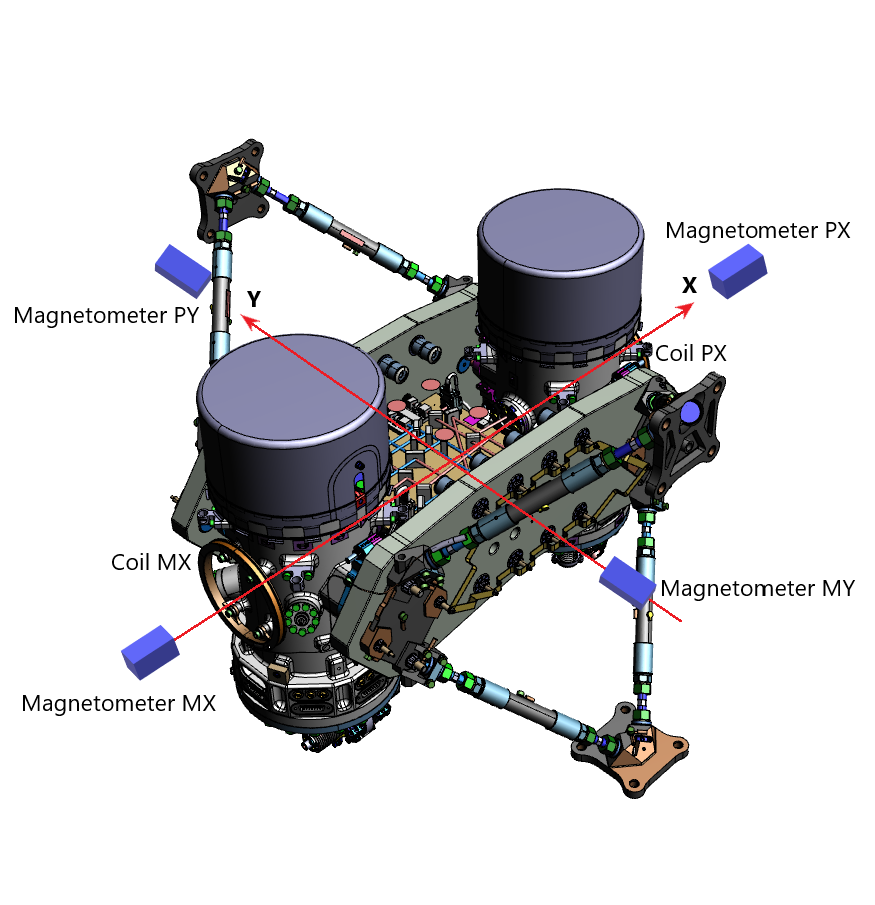} 
\caption{Magnetic elements of the LPF DDS: 2 coils and 4 tri-axial fluxgate magnetometers. The axis represents the spacecraft reference system frame.}
\label{fig.dds}
\end{center}
\end{figure}

The LISA Pathfinder Data and Diagnostics Subsystem (DDS) included a set of high precision sensors on-board precisely to monitor environmental disturbances with potential impact in the main scientific result of the mission, the relative acceleration between the free falling test masses. The DDS was composed by a temperature measurement subsystem~\citep{Sanjuan07, Armano19_Temp}, a magnetic diagnostic subsystem~\citep{Diaz-Aguilo13, Armano20_Mag}, see Fig.~\ref{fig.dds} for the distribution of the magnetic components, and a radiation monitor~\citep{Canizares09, Canizares11, Armano18_GCR1, Armano18_GCR2}. A crucial role of the DDS was to split up the experiment performance into different contributions to help on the design for future space-borne gravitational wave detectors. For that reason, the DDS also included a set of heaters and coils that were activated during flight operations to characterise the response of the instrument to intended environment disturbances.  
In this Letter we report on the results obtained in the estimate of the magnetic-induced contribution to the differential force noise of the test masses.

% FORCE NOISE CONTRIBUTION
\paragraph{Magnetically induced force noise contribution} 

Magnetic field fluctuations couple into the dynamics of the free falling test masses on board the satellite, which act as a dipole embedded in a magnetic field

\begin{eqnarray}
\mathbf{F} & = \left\langle \left( \mathbf{M_r} \cdot \mathbf{\nabla}\right)\mathbf{B} + \frac{\chi \rm{V}}{\mu_0}\,\left[\left(\mathbf{B} \cdot \mathbf{\nabla} \right)\mathbf{B} \right] \right\rangle,
\label{eq.dipole}
\end{eqnarray}

where in our notation bold letters stand for vectors, with $\mathbf{M_r}$ being the 3-dimensional intrinsic remanent moment of the magnetic dipole, $\chi$ its magnetic susceptibility, $\mathbf{B}$ the surrounding magnetic field and $\mu_{0}$ is the vacuum permeability. The right angle brackets denote TM volume, V, average of the enclosed quantity: $\left\langle \ldots\ \right\rangle \equiv \dfrac{1}{\rm{V}}\int_{\rm{V}} \,(\ldots)\,d^{3}x$. This expression can be further expanded~\citep{Diaz-Aguilo12}, however in this Letter we will be interested in the noise contributions mainly focusing on the x-component which is the most sensitive axis, connecting both TMs as defined by the spacecraft frame from Fig.~\ref{fig.dds}. Under reasonable assumptions --homogeneity and stationarity of the test mass properties-- we can express the main contributions to the acceleration noise budget as

\begin{eqnarray}
S_{F,x}(\omega) & = &  
\left( \frac{\chi V}{\mu_o}  \left< \mathbf{\nabla B_x}  \right>  \right)^2  S_\mathbf{B}(\omega) + \nonumber \\
& + &  \left( \mathbf{\left< M_r \right>}   + \frac{\chi V}{\mu_0}  \left< \mathbf{B_x}  \right>  \right)^{2}  S_{\mathbf{\nabla B_x}} (\omega)
 \label{eq.forceNoise} 
\end{eqnarray}

where $\rm S(\omega)$ stands for the Power Spectral Density (PSD). The first term in Eq.~(\ref{eq.forceNoise}) is the dominant one in the low frequency regime, since it couples fluctuations of the magnetic field ---with a $ f^{-1}$ behaviour due to the interplanetary contribution---  through an effective coupling $\sim \chi  \left| \mathbf{\nabla B_x}  \right| $. This coupling factor explains the stringent requirements set on the magnetic field gradient value generated by units on the spacecraft in the test mass location. Notice that while the magnetic susceptibility is a test mass property and, therefore, difficult to modify after the design phase for it to be as low as possible, reducing a potential gradient in the test mass location has an equal direct effect in the reduction of this leading term in the magnetic noise contribution. The second term in Eq.~(\ref{eq.forceNoise}) has a lesser impact in the low frequency band since the fluctuations associated with the gradient of the magnetic field show a flat spectrum, i.e. $\rm S_{\mathbf{\nabla B_x}} (\omega) \sim ct.$; we find this contribution to be orders of magnitude below.

% SETUP AND PARAMETERS
\paragraph{Determination of magnetic parameters} 
The magnetic diagnostic subsystem on-board LISA Pathfinder consisted of four tri-axial magnetometers and two induction coils. The two induction coils had a radius of 56.5\,mm and were built with 2400 windings of a copper alloy mounted on a titanium support, 85.5\,mm away from the test mass. Both coils were aligned with the axis joining both test masses so that the generated magnetic field had axial symmetry and driven by a dedicated high stability current driver to ensure that high precision magnetic forces were produced~\citep{Diaz-Aguilo13}. 
An extensive campaign of experiments were conducted to study magnetically-induced forces on the test masses during flight operations by applying controlled magnetic-induced forces in the test mass~\citep{Armano23_Mag_PRD}. The result of these suite of experiments can be summarised in two type of parameters as we show in Table~\ref{tbl.parameters}. A first set correspond to \emph{intrinsic} properties of the test mass, the magnetic susceptibility ($\chi$) and the remanent magnetic moment ($\mathbf{M_r}$) which obviously determine the reaction of the test mass to any applied magnetic field. A second set of parameters correspond to \emph{extrinsic} parameters to the test mass, the background magnetic field $\rm B_{\rm back., x} $  and magnetic field gradient $\rm \nabla_x B_{\rm back.,x} $, which also contribute to the test mass force when an external magnetic field is applied. 

  \begin{table}[t]
\begin{tabular}{cc}
Parameter & Value \\  
\hline
\hline
$\chi (\times 10^{-5})$  & $-3.3723 \pm 0.0069$  \\
$\rm M_{r, x}\, [nA m^{2}]$ & $0.140 \pm 0.138 $  \\
$\rm M_{r,y}\, [nA m^{2}]$ & $0.178 \pm 0.025 $  \\
$\rm M_{r,z}\, [nA m^{2}]$ & $0.095 \pm 0.010 $  \\
$\rm B_{\rm back.,x} [nT]$  & $414 \pm 74$  \\
$\rm \nabla_x B_{\rm back.,x} [nT m^{-1}]$  & $-7400 \pm 2100 $  \\
\end{tabular}
\caption{TM1 magnetic parameters as measured in-flight in~\citep{Armano23_Mag_PRD}. \label{tbl.parameters}} 
\end{table}

The magnetic susceptibility was estimated at different frequencies during in-flight experiments. We notice that the relevant impact in the noise contribution will be entering our estimates in the LISA Pathfinder band, i.e. in a low enough frequency regime where we can safely treat this parameter as a constant value since, for the LISA Pathfinder test masses, the susceptibility cut-off frequency is expected around the ~630\,Hz~\citep{Vitale07, Trougnou07}. Hence, for our calculations we will consider our best estimate, obtained at f = 5\,mHz.  
The value for the three components of the remanent magnetic moment are shown in Table~\ref{tbl.parameters}. All results were only measured for test mass \#1 due to the malfunctioning of the coil near the other test mass. We note here that, despite the high relative error carried by the measurement these are the most precise values obtained for the remanent moment of the test masses. The dedicated measurements on-ground could only find an upper limit of 4\,$\rm nA m^{2}$~\citep{tn_magneticmoment}, i.e. one order of magnitude above the ones presented here. 

\begin{figure*}[t]
\includegraphics[width=2\columnwidth]{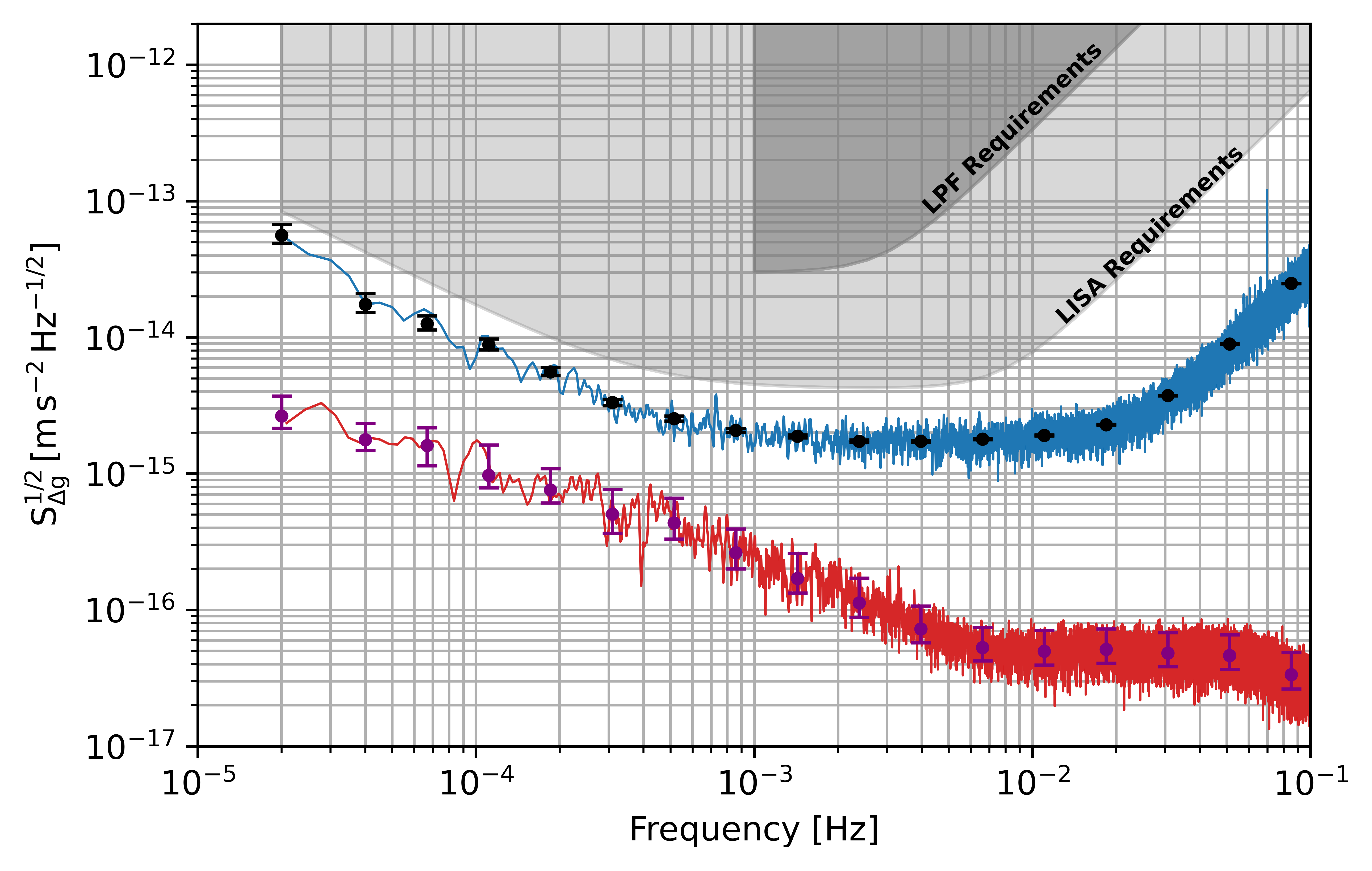}
\caption{Noise projection of the magnetic induced forces in the free-falling test masses relative acceleration. In blue the $S^{1/2}_{\Delta g}$ from Feb. 14th to Feb. 27th, 2017. In red the $\rm S^{1/2}_{\mathbf{B}}$ acceleration projection from the measurements of the magnetometers on-board during the same noise run. Black and purple dots represent an averaging of the spectra and the latter includes the uncertainty propagation after applying Eq.~\eqref{eq.forceNoise}.}
\label{fig.noiseProjection}
\end{figure*} 

When exerting a magnetic induced force with the coil we are able to recover the background magnetic field gradient, since it will add up to the magnetic field induced by the coil. 
The experiments on-board can only estimate the x-component for this contribution. The value obtained for TM1 during in-flight experiments, $\rm \nabla_x B_{\rm back.,x}  = (-7400 \pm 2100)\, nT/m$, points towards this
contribution originating in the thermistors attached to the external wall of the GRS Electrode Housing. These NTC (Negative Temperature Coefficient) sensors are manufactured using materials that can show ferromagnetic behaviour. Despite undergoing a degaussing process prior to their assembly in the satellite, their close proximity to the test mass ---roughly 13\,mm--- makes them prone to create a local gradient~\citep{Sanjuan08}. It is worth noting here that the estimates of $\rm B_{\rm back.,x}$ and $\rm \nabla_x B_{\rm back.,x}$ obtained during the experiments with coils on-board are the most precise attained in the test mass location, since magnetometers must be located sufficiently far from the test mass so that the magnetic field they generate do not perturb the test mass dynamics.

Given that the X-component of the local gradient is the only one measurable, we are forced to extrapolate these results to the Y and Z component and to the other test mass as well. To do so, a Monte Carlo (MC) simulation was performed with test mass \#1 and all eight NTCs surrounding it. For each NTC the remanence was randomly selected within the values [5$\times 10^{-7}$, 5$\times 10^{-5}$] Am$^{2}$, in accordance with the range of magnetic moments measured for these sensors during on-ground characterisation, and its orientation chosen at random in spherical coordinates. The magnetic field gradients across the volume of the test masses were found to be Gaussian distributed around zero with 5$\mu$T/m of standard deviation (for all components). To account that the LPF interferometric measurements involved the relative differences in positions of both  test masses, the MC uncertainties were multiplied by a factor of $\sqrt{2}$. Thus, the estimated total magnitude of the background magnetic field gradient was given mainly by $\nabla_x B_{\rm back.,x}$, as the other two components ($\nabla_y B_{\rm back.,x}$ and  $\nabla_z B_{\rm back.,x}$) obtained from the simulation were negligible, but its standard deviation took into account all MC results leading to $\rm\left|\mathbf{\nabla B_x} \right|  = 7.4_{-2.1}^{+5.5}\,\mu$T/m.

\begin{figure}[h]
     \centering
     \begin{subfigure}[b]{1\columnwidth}
         \centering
         \includegraphics[width=0.9\columnwidth]{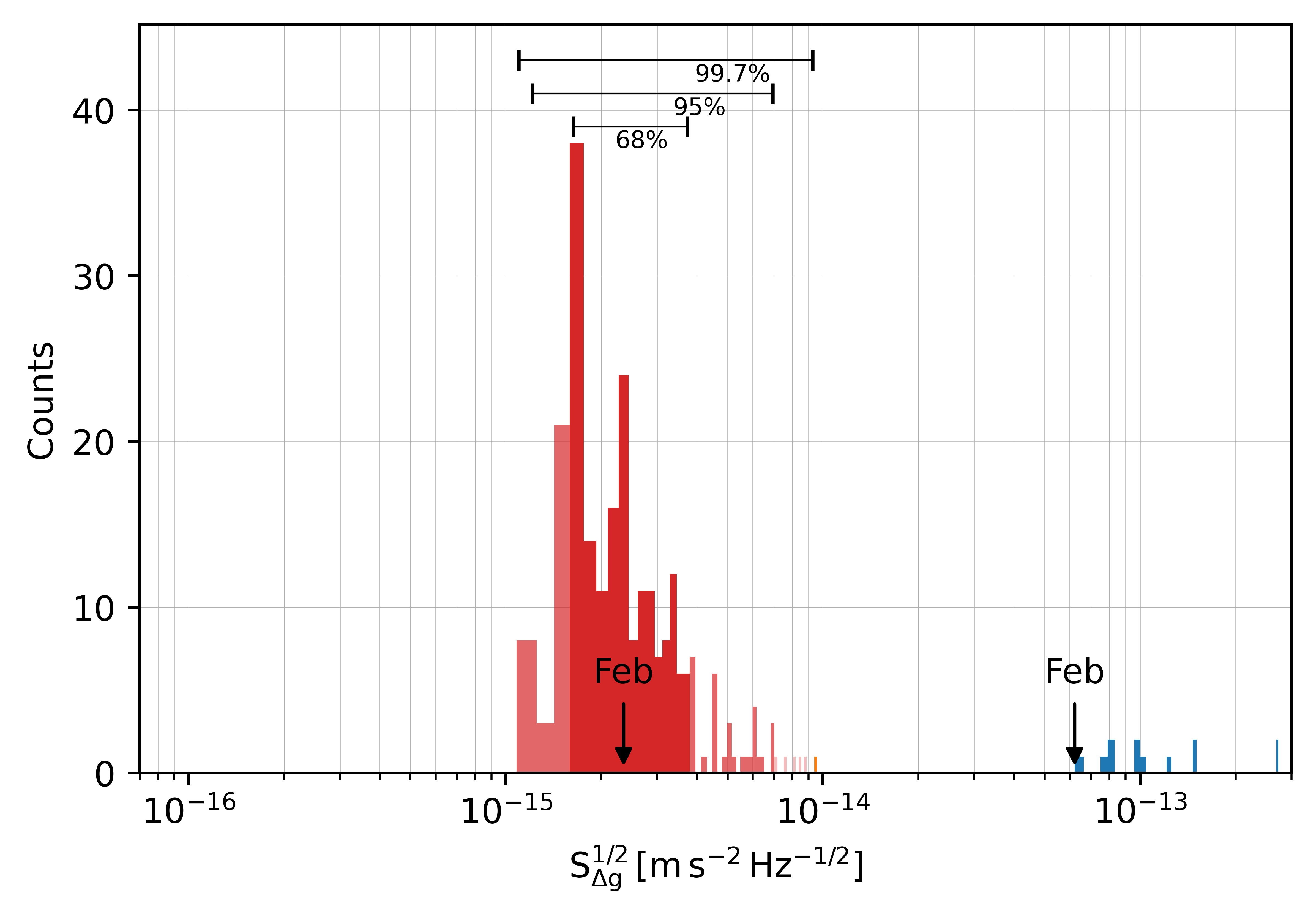}
         \label{fig.Magfluctuations.a}
     \end{subfigure}
     \begin{subfigure}[b]{1\columnwidth}
         \centering
         \includegraphics[width=0.9\columnwidth]{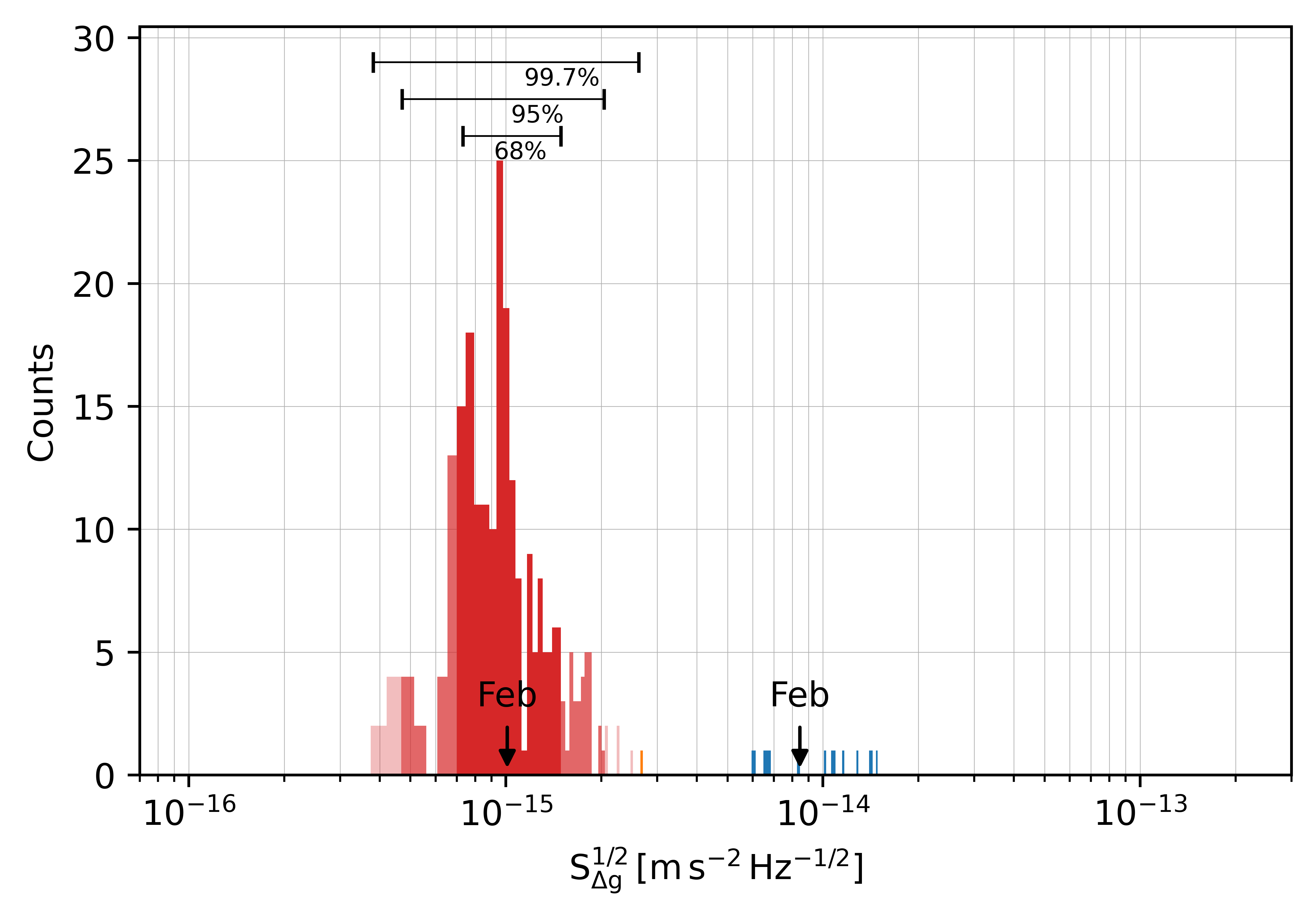}
         \label{fig.Magfluctuations.b}
     \end{subfigure}
     \begin{subfigure}[b]{1\columnwidth}
         \centering
         \includegraphics[width=0.9\columnwidth]{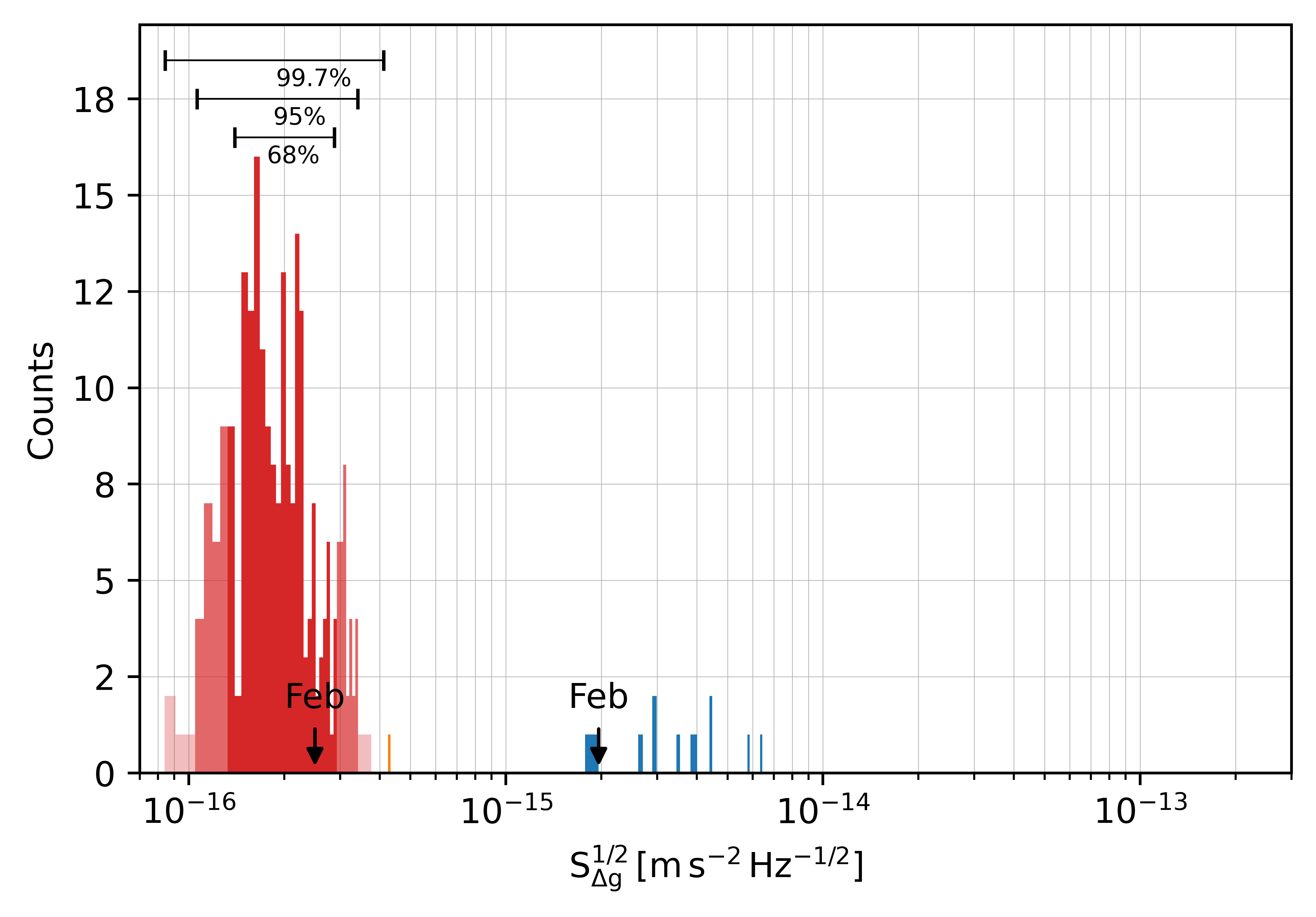}
         \label{fig.Magfluctuations.c}
     \end{subfigure}
        \caption{Statistical distribution of the magnetic contribution to acceleration noise (in red) at 20$\mu$Hz (top), 0.1 mHz (middle), and 1 mHz (bottom). The associated Gaussian distribution at 1$\sigma$, 2$\sigma$ and 3$\sigma$ ranges is also shown. For comparison, we display the median values of $S^{1/2}_{\Delta g}$ (in blue) for each of the 12 noise runs performed. The explicit values measured for both during the 2017 February run (Feb) are also shown.} \label{fig.Magfluctuations}
\end{figure}

% NOISE PROJECTION AND DISCUSSION
\paragraph{Results and Discussion} 
The main scientific result from LISA Pathfinder is expressed in terms of the differential force per unit mass acting on the two test masses, the so-called $\rm\Delta g$~\citep{Armano16}. In order to obtain this parameter, a series of signals were injected during the in-flight operations which allowed a complete description of the dynamic model of the motion of the test mass inside the spacecraft~\citep{Armano18_xtalk}. The model derived in this way was then used to translate the high precision interferometric measurement of the distance between the test masses into relative acceleration between them. The evaluation of this relative acceleration was performed in the so called noise runs, i.e. periods where the instrument was kept in its most stable, low-noise configuration for several days in order to obtain a good estimate of the $\Delta g$ spectrum down to the millihertz regime. 

In Fig.~\ref{fig.noiseProjection} we show the spectrum of fluctuations of the relative acceleration between both test masses in the noise run spanning from Feb. 14th at 01:59:50 to Feb. 27th at 09:53:29, 2017. The blue line represents the Amplitude Spectral Density (ASD) ---the square root of the PSD--- of the $\Delta g$ time series, $S^{1/2}_{\Delta g}$, which, in this segment, is well below the $\rm 30\,fm\,s^{-2}/\sqrt{Hz}$ at 1\, mHz required for LISA Pathfinder and even attaining the $\rm 3\, fm\,s^{-2}/\sqrt{Hz}$ at 1\, mHz required for LISA.  
The red line corresponds to the square root of only the first term in Eq.~\eqref{eq.forceNoise}. As previously explained, this is the leading contribution in the low frequency regime due to its coupling with the interplanetary magnetic fluctuations, $\rm S^{1/2}_{\mathbf{B}}$, and their $\propto f^{-1}$ dependence with frequency. 
Black dots with error bars represent the averaging of the spectra, $S^{1/2}_{\Delta g}$, at the same frequency bins used in~\citep{Armano18}. In purple, the same averaging is applied to the magnetic projection. In the latter case, however, the error bars include not only the statistical error from the Welch periodogram estimate but also the uncertainties from $\rm\left|\mathbf{\nabla B_x} \right|$, which includes the propagation of the MC analysis previously explained to Eq.~\eqref{eq.forceNoise}. According to our estimate, the magnetic contribution to $\Delta g$ during the February noise run accounts for a $\rm 1.46_{-0.77}^{+3.73}$\% in noise power at 0.1\,mHz.

We use the same data segment to evaluate the second term in Eq.~\eqref{eq.forceNoise}. These contributions couple to the $\Delta g$ through the fluctuations of the \emph{gradient} of the magnetic field. Being a gradient, fluctuations show a flat spectrum, reducing thus its relevance at low frequencies. We estimate its value to be $\rm \simeq 14\, am\,s^{-2}/\sqrt{Hz}$ which would start playing a significant role above 10\,mHz. However, from  Fig.~\ref{fig.noiseProjection} we can see that at around 7 mHz the spectrum of the first term flattens at $\rm \simeq 50\, am\,s^{-2}/\sqrt{Hz}$ instead of following the $f^{-1}$ tendency. This plateau is due to the intrinsic read-out noise of the magnetometers as it starts dominating the spectrum over the interplanetary magnetic fluctuations.

In Fig.~\ref{fig.noiseProjection} we have conveyed our complete understanding of the magnetic contribution to acceleration noise during the February 2017 noise run. However, the low frequency fluctuations of the interplanetary magnetic field show a non-stationary behaviour which follows closely the dynamics of the solar weather parameters, notably to the solar wind speed~\cite{Armano20_Mag}. 
In order to study this variability we extend our analysis to the complete magnetometers time series and, therefore, extrapolate the impact of this non-stationary behaviour in the estimation of the magnetic contribution to the relative acceleration between test masses for the whole mission duration. To do so, we compute the ASD of the interplanetary magnetic field at different frequencies logarithmically spaced within the very low frequency regime [20$\mu$Hz, 2 mHz] in segments of 8 days. This is to have more than 10 Welch periodogram averages at 20$\rm\mu$Hz, ensuring that the spectral window is not biasing our estimates while still having enough segments in total, 241, for the analysis to be statistically relevant. Afterwards, we project it into force noise by considering the first term in Eq.~\eqref{eq.forceNoise}.
In Fig.~\ref{fig.Magfluctuations} we show this magnetic projection ASD during the entire LPF mission in red by using the median value of $\rm\left|\mathbf{\nabla B_x} \right|$ and only considering the first term in Eq.~\eqref{eq.forceNoise}. We show this analysis for the frequency bins at 20$\mu$Hz, 0.1 mHz and 1 mHz, which correspond to the three panels in the figure. For comparison, we also display in blue the 12 noise runs performed during the whole mission.

Given that we have a sufficient statistics for the magnetic contribution to $\Delta g$, we can compute the associated confidence intervals at 1$\sigma$, 2$\sigma$ and 3$\sigma$ by considering the 16th and 84th percentiles, the 2.5th and 97.5th percentiles and the 0.15th and 99.85th percentiles, respectively. The distribution of the magnetic noise contribution implies that the $\rm 1.01_{-0.24}^{+0.73}\, fm\,s^{-2}/\sqrt{Hz}$ at 0.1 mHz that we have obtained for the February run conditions could increase to $\rm 2.70_{-0.77}^{+2.00}\, fm\,s^{-2}/\sqrt{Hz}$ for extreme solar wind conditions, where we have considered the uncertainty in $\rm\left|\mathbf{\nabla B_x} \right|$ in the error estimate.  

Finally, we want to emphasize that the the statistical distribution of the magnetic noise contribution due to the interplanetary magnetic field non-stationarity can not be directly associated with the variability of the measured acceleration noise in the test mass since the magnetic contribution is only one of the contributions --and not to dominant one-- to the overall $\Delta g$ model. Fig.~\ref{fig.Magfluctuations} compares this distribution to the few measured $S^{1/2}_{\Delta g}$ for the sake of completeness, although one should not derive a causal connection between both.

% NOISE PROJECTION AND DISCUSSION
\paragraph{Conclusions}
We have presented the first complete estimate of the magnetic-induced force noise contribution to the acceleration noise between free-falling test masses in the context of a space-borne gravitational wave detector. The results are based on a precise in-flight characterisation of those parameters affecting the magnetic behaviour of the test mass. 
Our results set a value for the magnetic contribution to the LISA Pathfinder acceleration noise during the noise run of February 2017 of $\rm 0.25_{-0.08}^{+0.15}\,fm\,s^{-2}/\sqrt{Hz}$ at 1\,mHz considering the magnetic gradient uncertainties, which is close to the initial assessments~\citep{Armano16} and well below the requirements established before launch for this contribution of  $\rm 12\,fm\,s^{-2}/\sqrt{Hz}$ at 1\,mHz. 
Our estimate at 0.1 mHz is $\rm 1.01_{-0.24}^{+0.73}\, fm\,s^{-2}/\sqrt{Hz}$, a $\rm 1.46_{-0.77}^{+3.73}$\% PSD contribution to $\Delta g$, which is in the order of other contributions such as the charging noise, with an estimated contribution of $\rm 1\, fm\,s^{-2}/\sqrt{Hz}$ at 0.1\,mHz~\citep{Armano17_charging} or the actuation noise, which is expected to be the dominant contribution with an expected value of $\rm 4.5\, fm\,s^{-2}/\sqrt{Hz}$~\citep{Armano17_capacitive} at 0.1 mHz.
We find that, as expected, the leading term of the magnetic-induced force noise to the test mass motion in these low frequencies originates from the coupling of fluctuations in the interplanetary magnetic field to the local magnetic field gradient through a constant term given by $\sim \chi  \left| \mathbf{\nabla B_x}  \right| $. 

Magnetic field fluctuations have a non-stationary behaviour related to space weather conditions, predominantly the solar wind speed, which will induce  excess test mass acceleration noise in the low frequency band for some extreme conditions of the solar wind. Those non-stationarities could rise the noise power contribution to $\Delta g$ of magnetic fluctuations up to a factor $\simeq 4.6$ at 0.1 mHz in the worst case scenario, that is during the most extreme solar wind condition, measured during LPF, and the largest predicted value of the gradient magnetic field $\rm\left|\mathbf{\nabla B_x} \right|  = 7.4_{-2.1}^{+5.5}\,\mu$T/m.
This low frequency magnetic field non-stationary behaviour is implicit to the interplanetary field and, hence, a potential disturbance for future space-borne gravitational wave observatories, although not being the dominant one in the case of LISA. 
The local magnetic field gradient in the test mass location is the design parameter that these future gravitational wave observatories in space can use to suppress this potential noise contribution. Our analysis shows that, in LISA Pathfinder, the temperature sensors (NTCs) located close to the test mass can be considered a potential source of the magnetic local gradient. Despite that it is not expected to limit the instrument performance in most of its operations, this potential excess noise contribution could be further suppressed by using Platinum temperature sensors with no magnetic contribution.
The results reported here are applicable to future space-borne detectors and other missions sharing the technology that was put to test by the LISA pathfinder mission.

%% ACKNOWLEDGEMENTS
\paragraph{Acknowledgements}
This work has been made possible by the LISA Pathfinder mission, 
which is part of the space-science program of the European Space Agency.
The French contribution has been supported by CNES (Accord Specific de 
projet CNES 1316634/CNRS 103747), the CNRS, 
the Observatoire de Paris and the University Paris-Diderot. 
E. P. and H. I. would also like to acknowledge the financial support of the UnivEarthS Labex program at Sorbonne Paris Cite\'e (ANR-10-LABX-0023 and ANR-11- IDEX-0005-02).
The Albert-Einstein-Institut acknowledges the 
support of the German Space Agency, DLR. 
The work is supported by the Federal Ministry for Economic Affairs and Energy based on a resolution of the German Bundestag (FKZ 50OQ0501 and FKZ 50OQ1601). The Italian contribution has been supported by 
Agenzia Spaziale Italiana and Instituto Nazionale di Fisica Nucleare. 
The Spanish contribution has been supported by Contracts No. AYA2010-15709 (MICINN),  No. ESP2013-47637-P, ESP2015-67234-P (MINECO); PID2019-106515GB-I00, PID2022-137674NB-I00 	(MICIU).
F. R. acknowledges support from a Formaci\'on de Personal Investigador (MINECO) contract. The Swiss contribution acknowledges the support of the Swiss Space Office (SSO) via the PRODEX Programme of ESA. L. F. acknowledges the support of the Swiss National
Science Foundation. The UK groups wish to acknowledge support from the
United Kingdom Space Agency (UKSA), the University of Glasgow, 
the University of Birmingham, Imperial College, and the Scottish Universities Physics Alliance (SUPA). J.I.T. and J.S. acknowledge the support of the U.S. National Aeronautics and Space Administration (NASA).